# Bimorphic Floquet Topological Insulators

Georgios G. Pyrialakos[1,3*], Julius Beck[2*], Matthias Heinrich[2],

Lukas J. Maczewsky[2], Nikolaos V. Kantartzis[3], Mercedeh Khajavikhan[4],

Alexander Szameit[2] and Demetrios N. Christodoulides[1]

**Topological theories have established a new set of rules that govern the transport properties in a wide variety of wave-mechanical settings. In a marked departure from the established approaches that induce Floquet topological phases by specifically tailored discrete coupling protocols[1-3] or helical lattice motions[4], we introduce a new class of bimorphic Floquet topological insulators that leverage connective chains with periodically modulated on-site potentials to unlock new topological features in the system. In exploring a "chain-driven" generalization of the archetypical Floquet honeycomb lattice, we identify a rich phase structure that can host multiple non-trivial topological phases associated simultaneously with both Chern-type and anomalous chiral states. Experiments carried out in photonic waveguide lattices reveal a unique and strongly confined helical edge state that, owing to its origin in bulk flat bands, can be set into motion in a topologically protected fashion, or halted at will, without compromising its adherence to individual lattice sites.**

[1] College of Optics & Photonics-CREOL, University of Central Florida, Orlando, Florida 32816, USA. [2] Institute for Physics, University of Rostock, Albert-Einstein-Str. 23, 18059 Rostock, Germany. [3] Department of Electrical and Computer Engineering, Aristotle University of Thessaloniki, GR-54124 Thessaloniki, Greece. [4] Ming Hsieh Department of Electrical and Computer Engineering, University of Southern California, Los Angeles, CA, USA. [*] These authors contributed equally to this work. [‡] email: demetri@creol.ucf.edu

Floquet engineering provides a powerful tool for shaping the topological structure of fermionic and bosonic settings alike[1-12]. At its core, it relies on the fact that a proper periodic modulation can induce chirality to arrangements of non-interacting particles, thereby giving rise to a wide range of topological insulators (TIs). Under these conditions, symmetry-protected states emerge along the boundaries of a modulated lattice, allowing for scatter-free transport on edges or interfaces with static (topologically trivial) domains. As such, helical transport constitutes a hallmark signature of complex topological order, and its presence in Floquet-driven systems highlights how periodic modulations can systematically extend the original classification of topological phases[13-19] beyond static systems with spin-orbit coupling or magnetic order. Along these lines, different topological invariants such as the winding number[20] are required to describe the topological nature of gaps with vanishing Chern numbers $\mathcal{C} = 0$ (Refs. [21-23]). These new degrees of freedom have enabled the experimental realization of a wide variety of topological systems in photonic lattices, ranging from anomalous Floquet TIs[1-3] to systems exhibiting Weyl point dynamics[24], Anderson TIs[25], TIs in synthetic dimensions[26] and photonic Z2 TIs exhibiting fermionic time reversal symmetry[27] to topological lasers[28]. Quite recently, anomalous driving protocols have enabled the observation of solitons in topological band gaps[29], the creation of nonlinearity-induced TIs[30] and investigations into the nonlinear dynamics of higher-order topological insulators[31].

Historically, photonic waveguide Floquet insulators can be divided into two distinct categories according to the character of the modulation involved. The first class is based on diatomic lattices (Fig. 1a), whereby a topological regime may be brought about by helical trajectories of the individual waveguide sites (Fig. 1b), introducing an effective chiral gauge field and, in turn, a non-trivial Chern insulating phase to the system[4]. On the other hand, more complex Floquet phases can be synthesized by step-wise coupling protocols that periodically allow for the selective interaction between neighboring sites (along different lattice vectors) by varying their corresponding separations (Fig. 1c) [1,2]. In this work, we introduce a third approach that embodies the distinctive characteristics of both aforementioned classes while overcoming their most pressing physical constrains. To this end, we introduce interstitial elements[32] between the sites of a static periodic lattice that act as Floquet drivers for the system. In doing so, these sites allow us to leverage the periodic modulation of their on-site potentials in order to synthesize hybrid topological lattices whose band structure simultaneously hosts both conventional Chern insulator bands ($\mathcal{C} = 1$) and anomalous topological bands ($\mathcal{C} = 0$), and is even capable of supporting topological phases with higher ranked invariants ($\mathcal{C} = 2$).

## Floquet driven potentials in a chained honeycomb lattice

To exemplify our approach, let us consider a honeycomb lattice of weakly-coupled elements, in which the three nearest-neighbor couplings vary independently in a cyclic fashion along the time (or propagation) coordinate. Despite a vanishing Chern number $\mathcal{C} = 0$, such arrangements are known to be capable of supporting different insulating and non-trivial Floquet topological phases[10]. The topological properties of this anomalous system are characterized by a three-dimensional winding number $\mathcal{W}$, which considers the temporal evolution of the system and counts the edge states that cross a particular band gap[20]. The Chern number of any band can be associated with the difference of the adjacent gap's winding numbers, $\mathcal{C} = \mathcal{W}_{above} - \mathcal{W}_{below}$ (Refs. [20,21]). It should be emphasized that a spatially fixed implementation of an anomalous FTI structure is impossible in a diatomic potential lattice. In other words, the accessible degrees of freedom, namely the potential magnitudes at sites A and B, cannot be individually controlled in any way that locally breaks time-reversal symmetry in the coupling Hamiltonian: The coupling terms (e.g. $c_1, c_2, c_3$ in a honeycomb lattice) would be mutually coherent. In addition, an uneven detuning between the on-site potentials A and B will inadvertently break the sublattice symmetry, and in turn introduce a trivial band gap that counteracts a topological transition.

To overcome these limitations, we introduce interstitial elements between the main sites of the potential lattice, one in each coupling path between adjacent sites of the two sublattices. Figure 2 illustrates how the band structure of the honeycomb system (Fig. 2a) is altered in the presence of these interstitial sites (Fig. 2b): Two copies of the honeycomb spectrum (Fig. 2a), each of which features a pair of Dirac cones in the first Brillouin zone, coexist symmetrically around multiple degenerate flat bands located at zero energy. The latter are comprised of compact localized states (CLS) that reside exclusively within the chain elements, locked by the destructive interference of light at the main sites. This particular modal structure is supported by the presence of additional symmetries in a system that hosts an odd total number of fundamental eigenmodes[33,34], here originating from the $5 \times 5$ bulk Hamiltonian

$$H(\mathbf{k}) = \sum_{j=1}^{3} \beta_j(t)\psi_j^\dagger \psi_j + \sum_{j=1}^{3} c_j(t)\left(\psi_j^\dagger \psi_A e^{\iota \mathbf{k}\cdot \boldsymbol{\delta}_{jA}} + \psi_j^\dagger \psi_B e^{\iota \mathbf{k}\cdot \boldsymbol{\delta}_{jB}}\right) + h.c. \tag{1}$$

where $\boldsymbol{\delta}_{jA}, \boldsymbol{\delta}_{jB}$ are lattice vector displacements between the $j$-th chain element and the main site of sublattice A and B, respectively. Out of the fifteen independent terms of the Hermitian Hamiltonian, only nine remain in equation (1), six involving nearest-neighbor interactions (with couplings $c_1, c_2, c_3$) and three involving on-site potential shifts exclusively at the interstitial sites ($\beta_1, \beta_2, \beta_3 \neq 0$, $\beta_A, \beta_B = 0$). Each of these terms can be addressed independently without perturbing the balance between sites A and B, thus fully preserving sublattice symmetry in the unit cell. In a continuous representation of the system (i.e. the Schrödinger equation with a continuous refractive index profile), the time-variation of both $c_j$ and $\beta_j$ can be realized via modification of the potential contrast at the interstitial sites – in other words, it is specifically these sites that can be leveraged to bring about a topological phase transition.

We consider how the system responds under sinusoidal modulations that are chirally phase-shifted by $2\pi/3$ between the three interstitial sites surrounding each primary site. In this case, the potential at the site between elements A and B of the $j$-th chain is given by $V_j(t) = 1 + \sin(2\pi t/T + 2\pi j/3)$, where $j \in \{1,2,3\}$ and $T$ denotes the Floquet period. Note that, by using an effective $2 \times 2$ tight-binding representation of the system (see Supplementary section S.I), this is equivalent to a continuous helical rotation of the principal direction of maximal coupling, despite the spatially static arrangement of the lattice sites. The topological changes due to the periodic drive can be traced by means of the unitary $\mathbf{k}$-space evolution operator $U(\mathbf{k}, t) = \mathcal{T}\exp\left(-i\int_0^t dt' H(\mathbf{k}, t')\right)$ where $\mathcal{T}$ denotes time-ordering. In this respect, we decompose $U = U_s U_d$ into a product of a quasi-static term $U_s = \exp(-iH_{\text{eff}} t)$ with $H_{\text{eff}} = i/T \log(U(\mathbf{k},T))$ and a dynamic term $U_d$ that accounts for the periodic micro-motion of the mode during the period of the drive with $U_d(\mathbf{k},T) = \mathbb{I}$ (Ref. [35]). Note that the effective Hamiltonian conforms to the traditional 10-fold way classification[15] which, in this two-dimensional setting, is characterized by a $\mathbb{Z}$ topological invariant. As depicted in the example of Fig. 2c, for a Floquet period of $T = 2\pi/7$ the quasi-energy spectrum $\varepsilon(\mathbf{k})$ of $H_{\text{eff}}$ exhibits $\mathcal{C} = 1$ in the upper- and lowermost bands, indicating the formation of a pair of Chern-type topological band gaps around the Dirac points. In turn, these gaps support helical edge states on the zig-zag edge similar to the ones in a conventional photonic FTI[4]. At the same time however, the action of the periodic chain drive opens another gap around $\varepsilon = 0$. Being nested between bulk bands with $\mathcal{C} = 0$, its anomalous topological nature is only revealed by the value of the winding number, i.e. $\mathcal{W} = 1$. While the topological edge states traversing this gap with constant slope are likewise helical, they inherit a key characteristic of the zero-energy flat band that they emerge from: The power in these states exclusively resides within the three interstitial sites of the edge unit cells (see Supplementary Fig. S3). Finally, the remaining degenerate flat band at $\varepsilon = 0$ shows that the quasi-energies of the chained lattice's bulk CLSs remain on average unaffected by the presence of the periodic modulation, i.e. the self-locking property of the flat-band modes clearly survives the topological phase transition. As a result, these bimorphic chain-driven lattices can host both compact localized states in the bulk and provide virtually dispersion-free mobility for tightly wave packets along at the edge.

## Observation of topological CLS states

In order to experimentally probe the propagation dynamics of the different topological states in chain-driven lattices, we employ femtosecond laser direct-written photonic lattices[36] as a platform for their implementation (see Methods). The evolution of light in such systems is governed by a Schrödinger equation in which the propagation coordinate $z$ represents time and the refractive index profiles of the individual waveguides act as interacting potential wells. In this context, the effective refractive index $\Delta n^{\text{eff}}$ of each waveguide provides direct control over the on-site potential and can be seamlessly tuned by modulating the inscription velocity along the propagation coordinate $z$. In turn, evanescent coupling between adjacent waveguides instantiates the required hopping terms. Having confirmed numerically that the desired characteristics of the $5 \times 5$ tight-binding model can be faithfully reproduced (see Supplementary section S.II) within the experimentally accessible parameter range of our platform, we fabricated triangular chain-driven lattices comprised of 42 unit cells (see Fig. 3a). Note that, despite its decidedly bristly appearance, the lattice is in fact terminated by the chained generalization of three zigzag-type edges, since the outermost waveguides represent interstitial sites. In a first set of experiments, we targeted the dispersive Chern states in the vicinity of the Dirac points by synthesizing a spectrally narrow wave packet of appropriate wave vector via a tripartite excitation pattern with alternating phases injected into three consecutive primary sites along the vertical edge of the system. A series of measurements for different initial positions (Fig. 3c) clearly shows a systematic counter-clockwise transport that is confirmed by extended-range numerical simulations (Fig. 3d) and allows light to circumnavigate the corners of the waveguide array. In contrast, when the alternating phase is removed from the excitation pattern, strong bulk diffraction was observed (Extended Data Fig. XD1), highlighting the absence of the Chern state in the center of the Brillouin zone ($k_{x,y} \approx 0$).

Owing to their flat-band origins and Brillouin-zone spanning nature (Fig. 4a), the anomalous topological edge states of the chain-driven lattice can be readily populated by injecting light into individual outermost interstitial sites. In a second set of experiments, we therefore traced the propagation of such single-site excitations along the edge and around two corners of the system (Fig. 4b), and observed helical topological transport in a counter-clockwise direction that, in contrast to the dispersive Chern channel, maintains the narrow width of the edge wave packet. This behavior is also confirmed by numerical extended-range propagation simulations (Fig. 4c). Single-site excitations of bulk interstitial sites instead remain localized at their initial positions (Fig. 4d,e) as dictated by the quasi-static part $U_q$ of the evolution operator, despite the fact that the dynamic part $U_d$ intermittently allows light to enter the neighboring sites during each Floquet cycle. Comparing these results to the discrete diffraction in a reference lattice of identical geometry implemented without the periodic modulation, we find that light injected into the interstitial sites on the edge as well as within the bulk of the non-driven lattice remains localized by virtue of the flat-band states residing there (Fig. 4f-j). This complementary behavior opens up the possibility to imprint arbitrary excitation patterns in the edge channels of chain-driven FTIs. The fact that the Floquet drive can be temporarily frozen and resumed at will without any changes to the lattice geometry provides unprecedented control over their topological transport dynamics: Once synthesized, compact wave packets can propagate along the edge without being subjected to dispersive broadening, while the unique properties of the flat band allows them to be freely shifted between travelling and localized states without ever rendering them vulnerable to bulk diffraction (see Supplementary Video).

## Discussion

Notably, the chained lattice exhibits a number of higher order topological phases that may occur for alternative periods of the drive. Changing the period $T$ of the modulation, in relation to the scale given by the inverse coupling in the system, allows adjacent unit cells of the $2\pi/T$-periodic quasienergy spectrum to overlap in a topologically non-trivial fashion. In the conventional coupling-modulated honeycomb system, a critical period $T_C = \pi/3$ separates the Chern phase ($\mathcal{C} = \pm 1$ and $\mathcal{W} = 0$ for $T > T_C$) and the anomalous phase ($\mathcal{C} = 0$ and $\mathcal{W} = 1$ for $T < T_C$), as illustrated in the Extended Data Fig. XD2. In contrast, the upper- and lowermost bands of the chain-driven lattice exhibit $|\mathcal{C}| > 1$ for any finite value of $T$, allowing for the Chern- and anomalous regimes to naturally coexist and interact in new and interesting ways: As $T$ is increased above a new critical point $T'_C = \pi > T_C$, the bands closer to zero energy enter a higher-order Chern phase ($\mathcal{C} = \pm 2$) as additional pairs of topological edge states emerge in their neighboring gaps (see Extended Data Fig. XD3).

In this work, we proposed and experimentally demonstrated a bimorphic class of Floquet topological insulators based on periodic modulations of certain on-site potentials. We showed how a "chained" honeycomb lattice, in which the exchange of population between primary sites is mediated by interstitial sites, can be endowed with rich topological features without resorting to magnetic interactions, helical lattice motion[4] or complex coupling protocols[1,2]. Beyond providing a complementary route towards inducing topology, our "chain- driven" systems synergistically combine the characteristic features of conventional and anomalous FTIs, allowing for the simultaneous existence of Chern-type chiral states and dispersion-less transport in quasi-localized wave packets. These anomalous modes bifurcate from the flat band of the static system, and, as such, can be readily converted into their likewise topologically protected, compact localized counterparts. While our findings are general and can be readily adapted to any topological platform that offers the means to dynamically control the on-site potential, such as cold atoms[37], electronic circuits[38] or even mechanics[39], the capability to affect topological phase transitions without changes to the lattice geometry is of particular importance in the context of topological photonics, where curved waveguide trajectories inevitably entail additional losses. Along these lines, we envision a new generation of low-loss robust photonic circuitry in which optically encoded packets of information can be transported, steered and even reshuffled without compromising their topological protection at any point.

# Figures

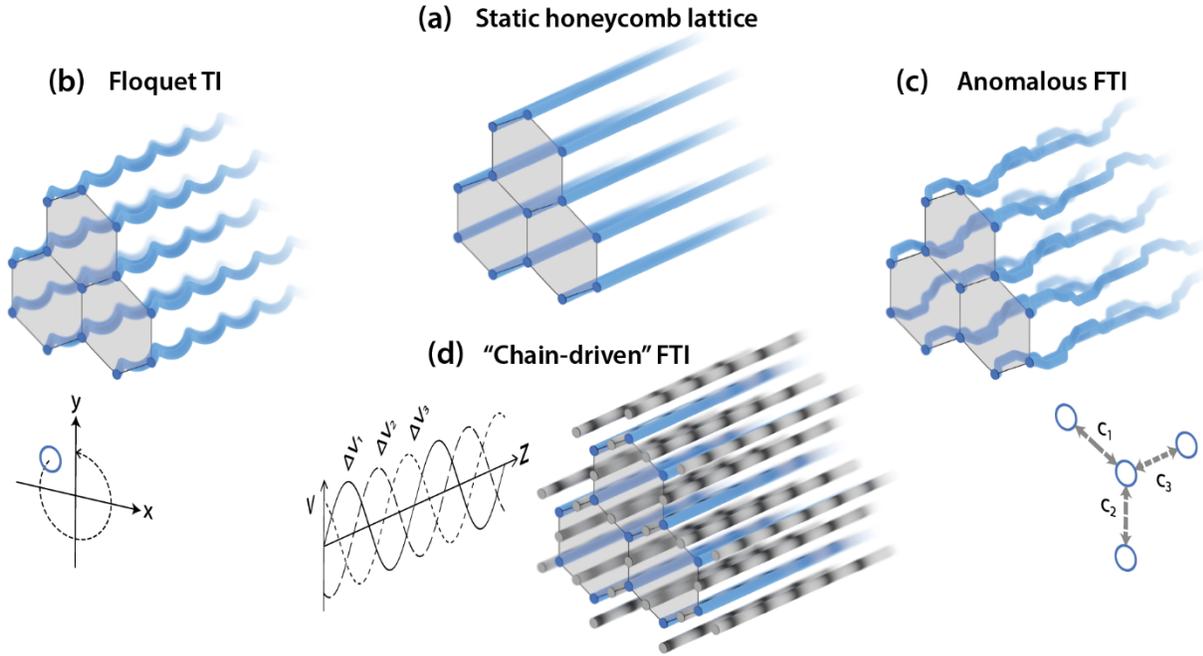

**Figure 1: A new road to topological lattices.** In the absence of magnetic interactions, periodic systems such as (**a**) the honeycomb lattice can be rendered topologically non-trivial by appropriate periodic modulations. (**b**) Conventional Floquet topological insulators achieve this by inducing a virtual magnetic flux via a global helical motion of the entire lattice, yielding a Chern-type topological phase. (**c**) Anomalous FTIs instead are based on multi-step driving protocols that impose helicity by independently modulating the coupling strengths between specific neighbors. Their topological properties are characterized by a winding number. Despite their different physical mechanisms, both of these approaches involve continual dynamic changes to the positions of the individual lattice sites – the main source of losses in topological photonics. (**d**) The "chain-driven" FTIs presented here instead leverage connective interstitial elements whose on-site potential is modulated in a cyclical fashion. Such systems offer rich topological phases that simultaneously support both Chern- and anomalous topological states in a geometrically static arrangement.

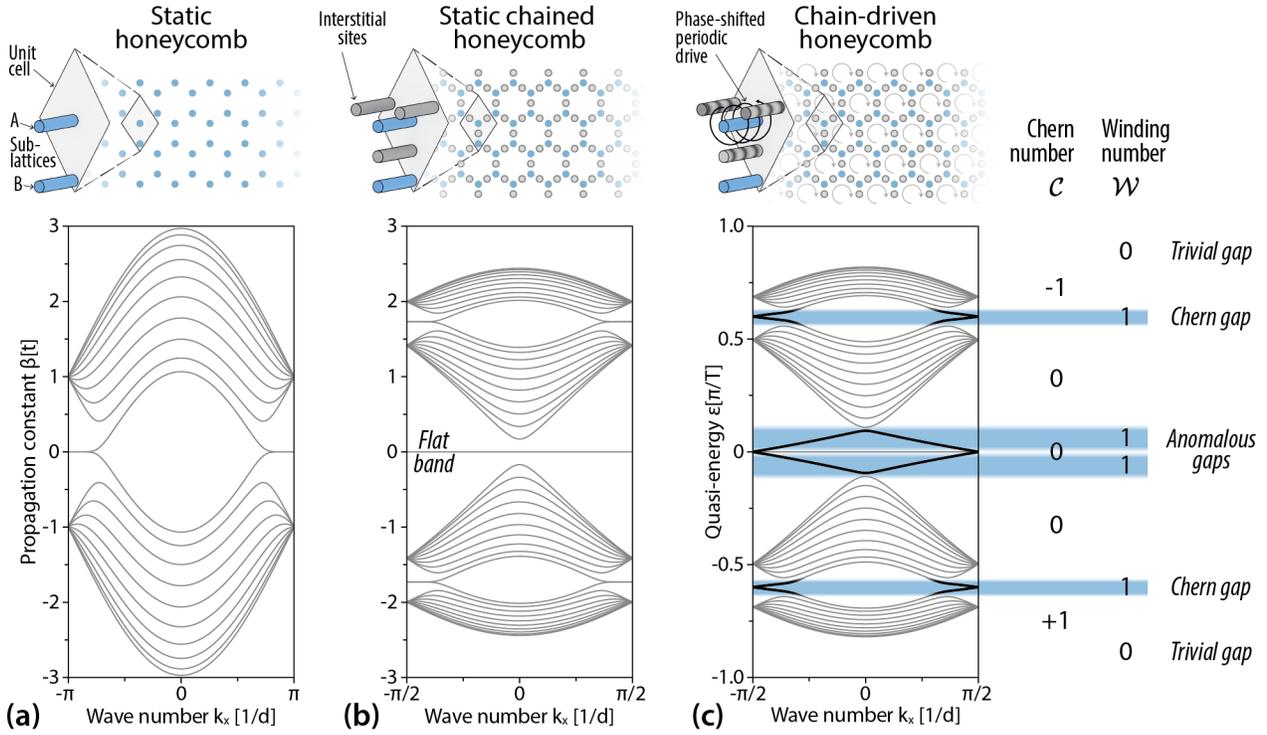

**Figure 2: Band structures and topological characterization.** (**a**) The zig-zag terminated edges of a conventional static honeycomb lattice support edge states that emerge from the Dirac Points and extend towards the boundaries of the first Brillouin zone. (**b**) A static chained honeycomb lattice is obtained by introducing by interstitial elements between each adjacent pair of principal sites. The resulting band structure manifests multiple degenerate flat-band states that are interposed between two copies of the diatomic spectrum. (**c**) A chain-driven honeycomb lattice. By modulating the on-site potential of the interstitial sites in a sinusoidal fashion with clockwise rotating relative phases, a total of four gaps open in the bulk, as indicated by the light blue shading. The edges support four pairs of helical states (thick lines), one for each topological gap. The values of the Chern ($\mathcal{C}$) and winding ($\mathcal{W}$) invariants, marked on the right, reveal the topological nature of each band gap. Notably, the edge states that emerge from the flat band resemble an anomalous topological phase, characterized by a non-trivial winding number ($\mathcal{W} = 1$) and a trivial Chern invariant 0 ($\mathcal{C} = 0$).

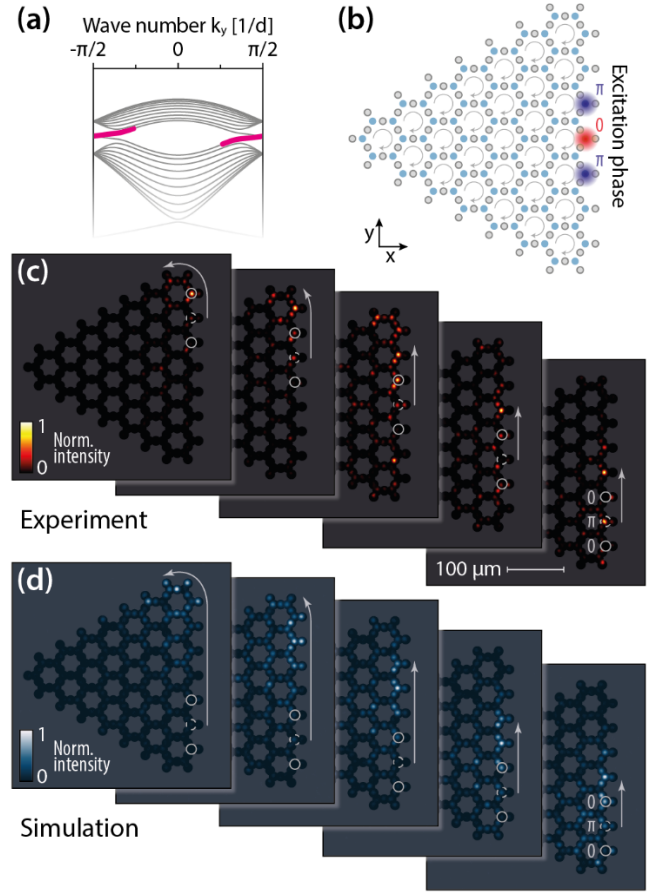

**Figure 3: Probing the Chern edge state of a photonic chain-driven honeycomb lattice**. (**a**) In order to selectively populate the topological state supported by the Chern gap (indicated by thick magenta branches in the ribbon band structure plot shown on the left), three consecutive primary sites along the zigzag edge are excited with identical amplitudes but alternating phases, synthesizing a spectrally narrow wave packet at the edge of the Brillouin zone. (**b**) Experimentally observed output intensity patterns at 633 nm after a propagation length of 150 mm for various placements of the initial excitation along the vertical edge of the lattice; positions and phase are shown by solid (phase 0) / dashed (phase $\pi$) white circles. As guide to the eye, the outlines of the respective lattice are indicated by a semi-transparent overlay. As confirmed by extended-range BPM simulations (**c**) starting from the lowest three unit cells, a substantial fraction of the launched light is captured by the topological Chern mode and transported along the edge in a counter-clockwise fashion. For comparison, the experimental and numerical results for equivalent flat-phase excitations are shown in Extended Data Fig. XD1: Without the staggered phase, light inevitably diffracts freely across the entire lattice.

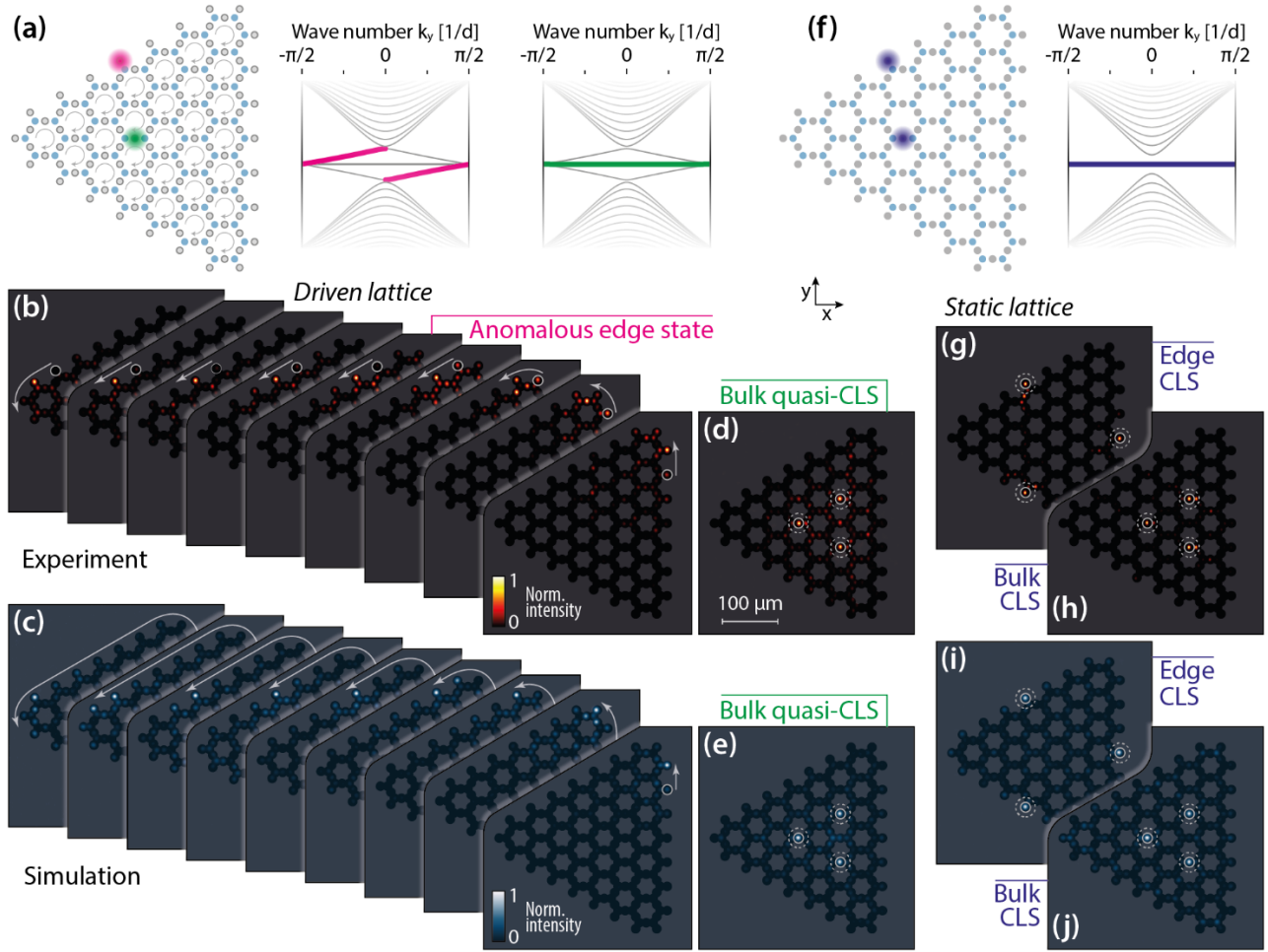

**Figure 4: Anomalous edge state propagation and compact localized bulk states. (a)** In contrast to the Chern modes, the flat-band-derived modes can be efficiently populated by single-waveguide excitations of the interstitial sites along the edge. **(b)** Experimentally observed output intensity patterns after a propagation length of 150 mm for various placements of the initial excitation along the edge of the chain-driven lattice. The injection positions are indicated by white circles. As guide to the eye, the outlines of the respective lattice are indicated by a semi-transparent overlay. As confirmed by extended-range BPM simulations (**c**), a substantial fraction of the launched light is captured by the anomalous topological mode and transported along the edge in counter-clockwise fashion. **(d,e)** Within the bulk of the chain-driven lattice, light injected into the interstitial sites remains trapped in the quasi-compact localized states and only undergoes a small degree of micro-motion during each Floquet period. **(f-j)** The degenerate flat band of the non-driven chained honeycomb lattice is comprised of compact localized states residing on the interstitial sites in the bulk as well as along the edge.

# Methods

**Experimental configuration.** The photonic structures used in our experiments are inscribed by focusing ultrashort laser pulses from a frequency-doubled fiber amplifier system (Coherent Monaco, wavelength 517 nm, repetition rate 333 kHz, pulse duration 270 fs) into the volume of a fused silica sample (Corning 7980, dimensions 1 mm × 20 mm × 150 mm, bulk refractive index $n_0 = 1.457$ at 633 nm), inducing permanent refractive index changes along arbitrary three-dimensional trajectories as defined by the motion of a precision translation system (Aerotech ALS130). Due to the focussing conditions, these waveguides exhibit slightly elliptical mode fields with a typical refractive index contrast of up to $\Delta n_0 = 2 \cdot 10^{-3}$. The selective modulation of the interstitial sites' index in a range of $\pm 10\%$ around this value was achieved by an appropriate modulation of the inscription speed between 92 and 156 mm/min. The propagation dynamics were probed with coherent light from a tunable supercontinuum source (NKT SuperK Extreme), allowing us to compensate for the micro-motion of the wave packets within the Floquet period and faithfully capture the dynamics according to the quasi-static evolution operator by varying the excitation wavelength between 570 nm and 633 nm. The appropriate intensity- and phase distributions for the desired excitation conditions were synthesized with a spatial light modulator (Hamamatsu LCOS-SLM).

**Numerical simulations.** The numerical results are obtained by solving the paraxial Schrödinger equation as an eigenvalue problem (for computations of the band structure – via the finite-difference method) and as a propagation problem (for computations of the field dynamics – via the beam-propagation method, BPM), respectively. In this context, the ribbon band diagrams of Fig. 2 and Fig. 3 provide the necessary validation for the tight-binding chain approximation.

To efficiently reduce the computational burden of the time-dependent eigenvalue problem, we decompose the solution space into a sinusoidal basis along the propagation axis. This approach lies on the expectation that the time-periodic part of the eigenmode solution will be related to the driving protocol used for the time-dependent modulation of the refractive index. This strategy leads into a solution space that is numerically large in the transverse plane (discretized by finite differences) but highly reduced along the $z$ dimension. The size of the matrix generated through this process is optimally minimized so that it remains within reach of common eigenvalue decomposing techniques. For more information see Supplementary section S.II.

# Data availability

The experimental data that support the findings of this study are available from M.H. upon reasonable request.

# Code availability

The MATLAB® codes corresponding to the BPM and band structure algorithms are available from G.G.P. upon reasonable request.

## Author contributions

G.G.P. and J.B. contributed equally to this work. G.G.P initiated the idea, formulated the index-modulated lattice and performed the theoretical calculations and simulations. J.B. developed the experimental implementation, fabricated the samples and conducted the measurements. J.B., L.J.M and M.H. evaluated the measurements and interpreted the data. M.K., N.V.K., A.S. and D.N.C. supervised the efforts of their respective groups. All authors discussed the results and co-wrote the manuscript.

## Acknowledgements

The authors would like to thank C. Otto for preparing the high-quality fused silica samples used for the inscription of all photonic structures employed in the experiments presented here. G.G.P. acknowledges the support of the Bodossaki Foundation. This work was partially supported by DARPA (D18AP00058), ONR MURI (N00014-16-1-2640, N00014-18-1-2347, N00014-19-1-2052, N00014-20-1-2522, N00014-20-1-2789), AFOSR MURI (FA9550-20-1-0322, FA9550-21-1-0202), National Science Foundation (NSF) (DMR-1420620, EECS-1711230, ECCS-1454531, DMR-1420620, ECCS-1757025, CBET-1805200, ECCS-2000538, ECCS-2011171), MPS Simons collaboration (Simons grant 733682), W. M. Keck Foundation, US–Israel Binational Science Foundation (BSF: 2016381), US Air Force Research Laboratory (FA9550-14-1-0037, FA9550-20-1-0322, FA9550-21-1-0202, FA86511820019,). The authors furthermore acknowledge funding from the Deutsche Forschungsgemeinschaft (SCHE 612/6-1, SZ 276/12-1, BL 574/13-1, SZ 276/15-1, SZ 276/20-1) and the Alfried Krupp von Bohlen and Halbach foundation.

## Competing interests

The authors declare no competing financial interests.

## Materials & correspondence

Correspondence and requests for additional materials should be addressed to D.N.C.

# Extended Data Figures

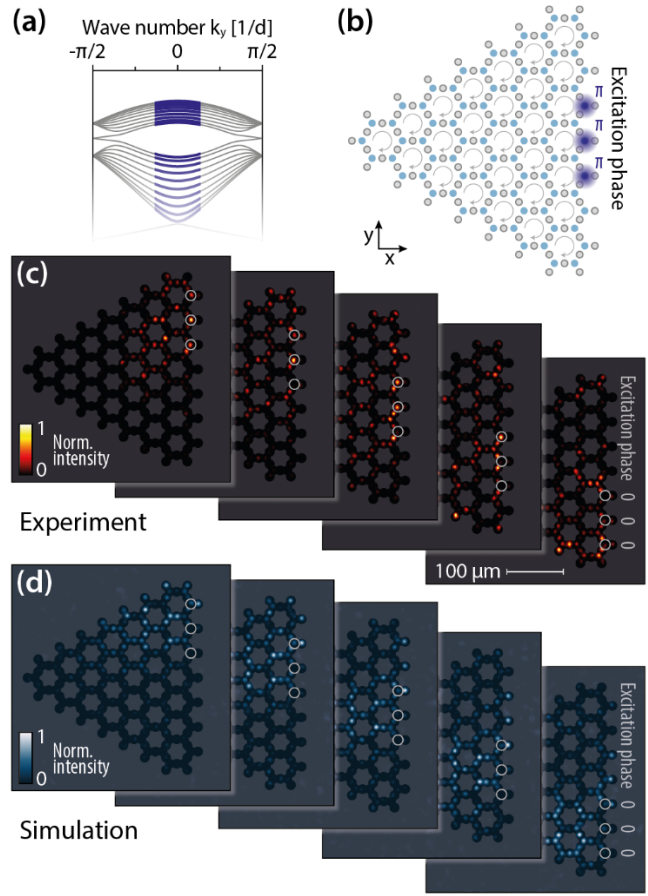

**Extended Data Figure XD1: Broad flat-phase excitation**. Absent the staggered phase, three-waveguide excitations of the primary sites along the edge fail to address the Chern mode and instead populate the bulk bands near the center of the Brillouin zone and instead place the wave packet symmetrically at its center. The resulting population of the bulk bands allows the light to diffract freely across the entire lattice. The arrangement of panels corresponds to Figure 3 of the main manuscript.

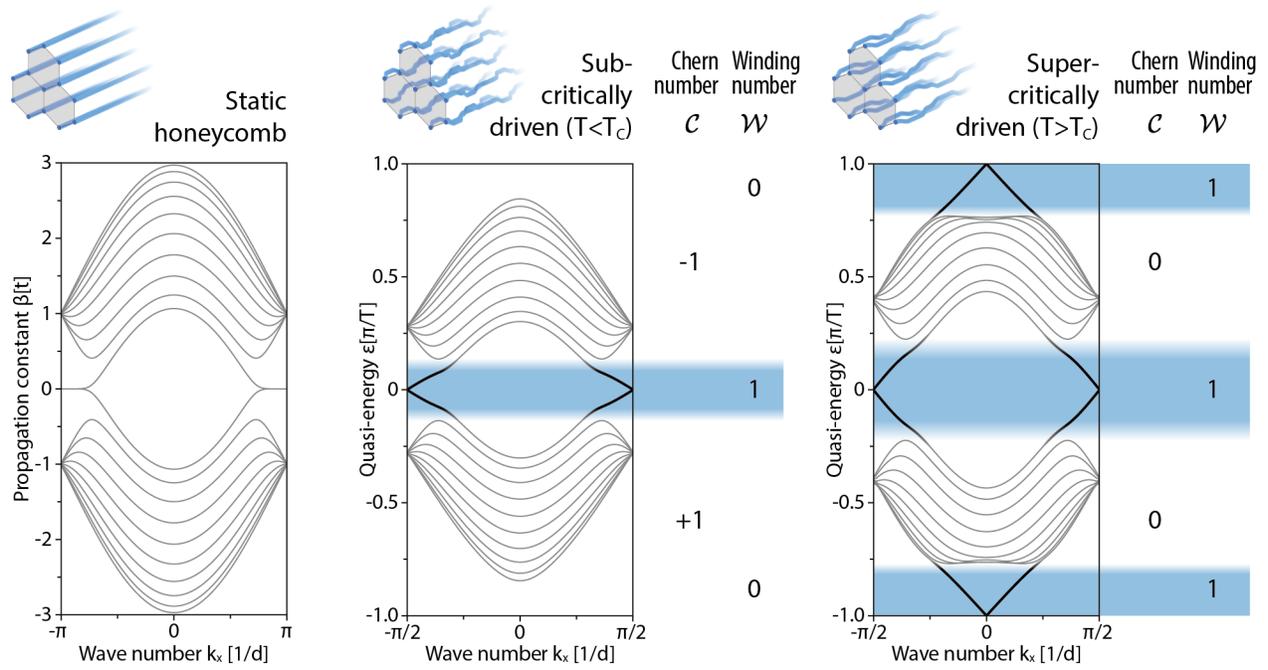

**Extended Data Figure XD2: Honeycomb helical FTI.** A driven honeycomb lattice with sinusoidally modulated time-periodic coupling terms exhibits a secondary topological phase in response to an increase of the driving period. At the critical driving period $T_C = \pi/3$ the gap at the Floquet zone collapses and reopens with a topologically non-trivial winding number. This corresponds to an anomalous phase with a trivial Chern number, signifying a topological phase transition.

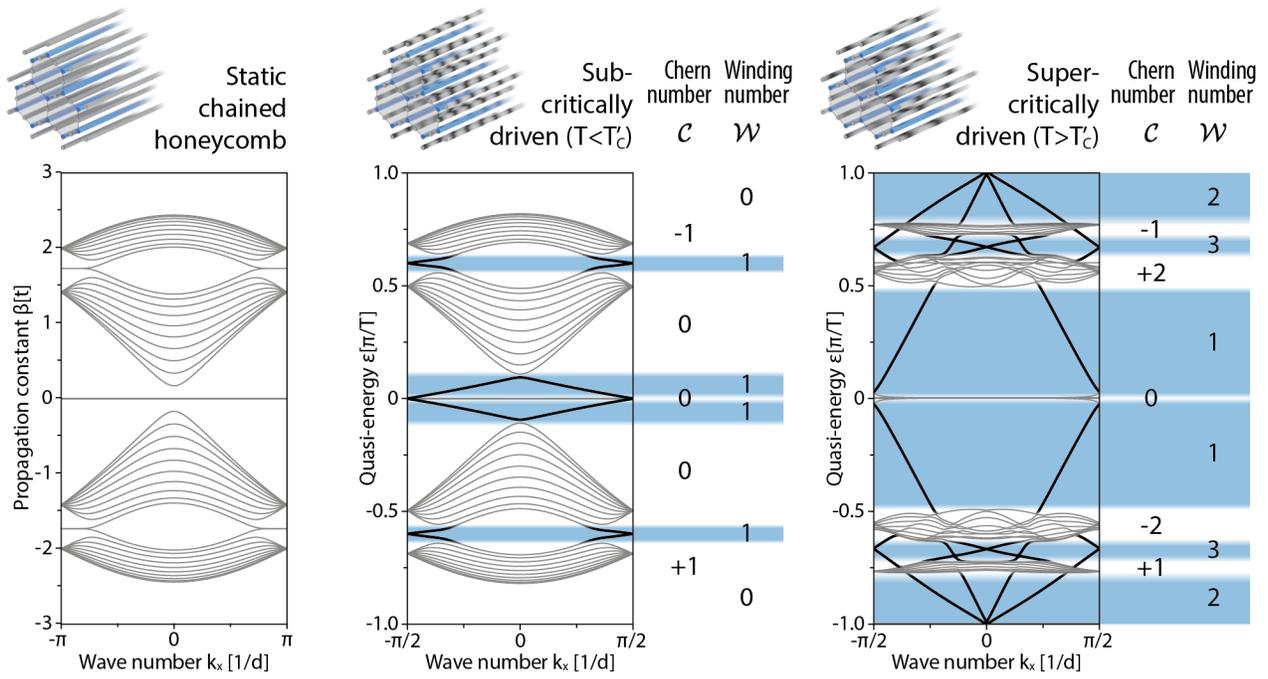

**Extended Data Figure XD3: Chained honeycomb FTI.** Above its critical modulation period $T'_C = \pi$, the chain-driven honeycomb lattice enters a secondary topological phase characterized by the band diagram shown on the right. In this configuration, the bands manifest a non-trivial topological structure characterized by higher order Chern invariants ($\mathcal{C} = 2$), and, in turn, the Chern gaps host an increased number of unidirectional edge states.